%% file: Hybrid_data.tex
\documentclass[12pt]{iopart}

\usepackage{latexsym}
\usepackage{graphicx}
\usepackage[usenames]{color}

\input epsf

\newcommand{\eqref}[1]{(\ref{#1})}
 
\newcommand{\MAH}{M_{\rm AH}}
\newcommand{\Mirr}{M_{\rm irr}}
\newcommand{\Rext}{R_{\rm ext}}

\input{definitions}

\begin{document}

\title[Hybrid black-hole binary initial data]
{Hybrid black-hole binary initial data}

\author{Bruno C. Mundim$^1$, Bernard J. Kelly$^{2,3}$, 
Yosef Zlochower$^1$, Hiroyuki Nakano$^1$, Manuela Campanelli$^1$}

\address{$^1$ Center for Computational Relativity and Gravitation, 
School of Mathematical Sciences, 
Rochester Institute of Technology, Rochester, New York 14623, USA}

\address{$^2$ CRESST \& Gravitational Astrophysics Laboratory, NASA/GSFC,
8800 Greenbelt Rd., Greenbelt, MD 20771, USA}
\address{$^3$ Dept. of Physics, University of Maryland, Baltimore County,
1000 Hilltop Circle, Baltimore, MD 21250, USA}

\ead{\mailto{bcmsma@astro.rit.edu}, \mailto{bernard.j.kelly@nasa.gov}, 
\mailto{yosef@astro.rit.edu}, \mailto{nakano@astro.rit.edu}, 
\mailto{manuela@astro.rit.edu}}

\begin{abstract}
Traditional black-hole binary puncture initial data is conformally flat. 
This unphysical assumption is coupled with a lack of radiation 
signature from the binary's past life. As a result, waveforms 
extracted from evolutions of this data display an abrupt jump.
In Kelly \etal~\cite{Kelly:2009js}, a new binary black-hole initial data 
with radiation contents derived in the post-Newtonian (PN) calculation 
was adapted to puncture evolutions in numerical relativity. 
This data satisfies the constraint equations to the 2.5PN order, 
and contains a transverse-traceless ``wavy'' metric contribution, 
violating the standard assumption of conformal flatness. 
Although the evolution contained less spurious radiation, 
there were undesired features; 
the unphysical horizon mass loss and the large initial orbital eccentricity. 
Introducing a hybrid approach to the initial
data evaluation, we significantly reduce these undesired features. 
\end{abstract}

\submitto{\CQG}

\maketitle

\section{Introduction}

The field of Numerical Relativity (NR) has progressed at a remarkable
pace since the breakthroughs of 2005~\cite{Pretorius:2005gq,
Campanelli:2005dd, Baker:2005vv} with the first successful fully
non-linear dynamical numerical simulation of the inspiral, merger, and
ringdown of an orbiting black-hole binary (BHB) system.  In
particular, the ``moving-punctures'' approach, developed independently
by the NR groups at NASA/GSFC and at RIT, has now become the most
widely used method in the field and has been successfully applied
to evolve generic BHBs.  This approach regularizes a singular
term in the space-time metric and allows the black holes (BHs) to
move across the
computational domain. Since this breakthrough, BHB physics has rapidly matured into a critical tool for
gravitational wave (GW) data analysis and astrophysics.  Recent
developments include: studies of the orbital dynamics of spinning
BHBs~\cite{Campanelli:2006uy, Campanelli:2006fg, Campanelli:2006fy,
Herrmann:2007ex, Marronetti:2007ya, Marronetti:2007wz, Berti:2007fi},
calculations of recoil velocities from the merger of unequal-mass
BHBs~\cite{Herrmann:2006ks, Baker:2006vn, Gonzalez:2006md}, the
surprising discovery that very large recoils can be
acquired by the remnant of the merger of two spinning BHs
~\cite{Herrmann:2007ac,
Campanelli:2007ew, Campanelli:2007cg, Lousto:2008dn, Pollney:2007ss,
Gonzalez:2007hi, Brugmann:2007zj, Choi:2007eu, Baker:2007gi,
Schnittman:2007ij, Baker:2008md, Healy:2008js, Herrmann:2007zz,
Herrmann:2007ex, Tichy:2007hk, Koppitz:2007ev, Miller:2008en},
empirical models relating the final mass and spin of
the remnant with the spins of the individual BHs
~\cite{Boyle:2007sz, Boyle:2007ru, Buonanno:2007sv, Tichy:2008du,
Kesden:2008ga, Barausse:2009uz, Rezzolla:2008sd, Lousto:2009mf}, and
comparisons of waveforms and orbital dynamics of
BHB inspirals with post-Newtonian (PN)
predictions~\cite{Buonanno:2006ui, Baker:2006ha, Pan:2007nw,
Buonanno:2007pf, Hannam:2007ik, Hannam:2007wf, Gopakumar:2007vh,
Hinder:2008kv, Campanelli:2008nk}.

One of the important applications of NR is the generation of waveforms
to assist GW astronomers in their search and analysis of GWs from the
data collected by ground-based interferometers, such as
LIGO~\cite{Abbott:2007kv} and VIRGO~\cite{2008CQGra..25k4045A}, 
and future missions, such as LCGT~\cite{2010CQGra..27h4004K}, 
LISA~\cite{LISA1}, ET~\cite{2010CQGra..27s4002P} and 
DECIGO~\cite{Ando:2010zza}. BHBs are particularly
promising sources, with  the final merger event producing a strong
burst of GWs at a luminosity of $L_{GW}\sim 10^{22}L_{\odot}$\footnote{
This luminosity estimate is independent of the binary mass and takes
into account that $3-10\%$ of the total mass $M$
of the binary is radiated over a time interval of $\sim100M$
\cite{Lousto:2009mf}.}, greater
than the combined luminosity of all stars in the observable universe.
The central goal of the field has been to develop the theoretical
techniques, and perform the numerical simulations, needed to explore
the highly dynamical regions and thus generate GW signals from a
representative sample of the full BHB parameter space. Accurate
waveforms are important to extract physical information about the
binary system, such as the masses of the components, BH spins, and
orientation.

With this in mind, we note a drawback shared by most present-day comparable-mass
black-hole binary simulations in three spatial dimensions: they are performed using
conformally flat initial data given by the Bowen-York (BY)~\cite{Bowen:1980yu} prescription as applied by
Brandt and Br\"{u}gmann~\cite{Brandt:1997tf}. This prescription, while numerically
convenient, lacks physical realism. The unique stationary vacuum black-hole solution
to Einstein's equations is the Kerr solution, which is not conformally flat for non-zero spin. We cannot
approximate this with single-puncture spinning BY data without also including unphysical
radiation.

Conversely, approximating an inspiralling black-hole binary system with two-puncture
BY data will leave out the gravitational radiation expected in physical situations --
radiation inextricably linked to the past history that produced the inspiral. This is
demonstrated in plots of extracted radiation from current simulations, where observers
a distance $\Rext$ from the binary only see a flat radiation profile for the first
$t \approx \Rext$ of evolution time.

The post-Newtonian (PN) initial data developed in this and preceding 
papers~\cite{Tichy:2002ec,Kelly:2007uc,Kelly:2009js} is an
attempt to address this shortcoming in initial data by incorporating to leading PN accuracy the gravitational-wave
content of a physical binary inspiral. The previous paper in this series, \cite{Kelly:2007uc},
demonstrated the evolution behaviour of the data evolved with no numerical conditioning of
the constraints. In summary, we noted that the extracted radiation agreed with physical expectations
from the very start of the simulation (though the burst of junk radiation associated with
puncture evolutions was not completely removed). We also encountered several related weaknesses
in the new evolved data, including:
\begin{itemize}
\item very high eccentricity ($\sim 10\%$) in puncture trajectories, until around $100M$ before merger;
\item (possibly related) an extremely slow stabilization of pre-merger horizon masses during evolution;
\item large constraint violations when all the gravitational-wave terms are included.
\end{itemize}
In this paper, we present hybrid BY-PN data with partial constraint conditioning, aimed at resolving
these issues.

In Section~\ref{sec:BY_ID}, we summarize the standard conformally flat Bowen-York puncture initial data.
In Section~\ref{sec:PN_ID}, we review the theoretical work that led to the PN data, and
discuss the encoding of the past history of the binary in Section~\ref{sec:PNtraj}.
In Section~\ref{sec:HID}, we present the new hybrid initial data, and demonstrate its improved
numerical behaviour. 
The numerical evolutions and comparison for various initial data 
are discussed in Section~\ref{sec:num}. 
We conclude in Section~\ref{sec:CON} with a discussion of future development
and application of this data. 

%%%%%%%%%%%%%%%%%%%%%%%%%%%%%%%%%%%%%%%%%%%%%%%%%%%%%%%%%%%%%%%%%%%%%%%%
\section{BY Initial Data}\label{sec:BY_ID}
%%%%%%%%%%%%%%%%%%%%%%%%%%%%%%%%%%%%%%%%%%%%%%%%%%%%%%%%%%%%%%%%%%%%%%%%

Most groups use the puncture prescription of Brandt and Br\"{u}gmann~\cite{Brandt:1997tf}
with Bowen-York extrinsic curvature~\cite{Bowen:1980yu} to initialize the numerical fields
for a puncture evolution.
In this prescription, the three-metric $\gamma_{ij}$ is conformally flat:
\be
\gamma_{ij} = \psi^4 \de_{ij} \,, \label{eq:gij_conformal}
\ee
and the conformal factor $\psi$ must satisfy the Hamiltonian constraint
\be
\Delta \psi + \frac{1}{8} K^{ij}K_{ij} \psi^{-7} = 0 \,, \label{eq:CHam_conformal}
\ee
where the conformal Bowen-York extrinsic curvature $K_{ij}$ already satisfies the momentum
constraint for holes with arbitrary momentum and spin.
  Brandt and Br\"{u}gmann's insight to find the solution to equation~(\ref{eq:CHam_conformal})
was to factor out the divergent parts of $\psi$ , leaving a well-behaved,
simply-connected sheet on which to solve their modified constraint:
\beq
\psi_{\rm BY} = 1 + \sum_{a=1}^2 \fr{m_a}{2r_a} + u \,,
\eeq
where BY refers hereafter to Bowen-York puncture data, 
$r_a=|{\bf x}-{\bf x}_a|$, and $m_a$ and ${\bf x}_a$ denote 
the ``bare'' mass and location of each hole, respectively. 
  In this prescription, only a single elliptic equation for $u$ 
has to be solved, and $u$ will be regular everywhere on the grid.
  Despite the great success in simplifying the form of the constraint
equations, this prescription does \emph{not} make use of any information
to accurately describe the past evolution of the binary black holes.
Conformal flatness, for example, prevents the astrophysically expected
gravitational radiation from being included in the initial data.

%%%%%%%%%%%%%%%%%%%%%%%%%%%%%%%%%%%%%%%%%%%%%%%%%%%%%%%%%%%%%%%%%%%%%%%%
\section{PN Initial Data in ADM-TT Gauge}\label{sec:PN_ID}
%%%%%%%%%%%%%%%%%%%%%%%%%%%%%%%%%%%%%%%%%%%%%%%%%%%%%%%%%%%%%%%%%%%%%%%%

Post-Newtonian (PN) techniques are considered to \emph{accurately}
represent an astrophysical system in the limit of slow motion/far-apart black holes.
See Blanchet~\cite{Blanchet:2006LR} and Sch\"afer~\cite{Schaefer:2009dq}
for a review of the PN approach.

In the canonical Hamiltonian formulation with point-like sources in general relativity,
Ohta \etal~\cite{Ohta:1973qi,Ohta:1974uu,Ohta:1974kp} 
investigated a class of coordinate systems in which the metric tensor becomes 
Minkowskian at spatial infinity, and developed the
2PN order calculation.
In particular, they found a compact way of expressing a compatible
coordinate condition with the aid of the ADM formalism.
By applying a transverse-traceless decomposition of the three-metric and its
conjugate momentum, they were able to encode the full information about
the dynamics of the canonical fields and particle variables in a reduced
Hamiltonian, which is obtained by solving the constraint equations
\emph{only}.
To solve the constraints, we follow the steps of Sch\"afer~\cite{Schaefer85}. 
Historically, the 2PN order calculation 
was completed by Damour and Sch\"afer~\cite{Damour1985}. 
Shortly afterward, Sch\"afer~\cite{Schaefer85,Schaefer86} included
the 2.5PN radiation-reaction terms in this Hamiltonian approach.

Tichy \etal~\cite{Tichy:2002ec} adapted the 2.5PN ADM-TT results to puncture initial
data for numerical relativity.
We start then from the PN expression for the spatial metric, which differs from
conformally flatness by a radiative term $h^{\rm TT}_{ij}$:
\beq
\ga^{\rm PN}_{ij}=\psi^4_{\rm PN} \de_{ij} + h^{\rm TT}_{ij} \,, \label{eq:gapn}
\eeq
where the PN conformal factor is given by
\beq \label{eq:psiPN}
\psi_{\rm PN}=1+\sum_{a=1}^2 \fr{E_a}{2r_a} + O(\ep^6) \,,
\eeq
where $\ep \equiv 1/c$ is the PN order parameter and
\beq
E_a=\ep^2 m_a + \ep^4 \lb \fr{{\bf p}^2_a}{2m_a} - \fr{m_1m_2}{2r_{12}} \rb \,.
\eeq
Here, ${\bf p}_a$ represents the linear momentum of each hole,
and $r_{12}$ is the separation between the holes. 

The accompanying extrinsic curvature $K^{\rm PN}_{ij}$ is related to the three-metric's
(trace-free) conjugate momentum $\pi_{\rm PN}^{ij}$:
\beq
\pi_{\rm PN}^{ij} = \fr{1}{\psi_{\rm PN}^4} 
\lb \ep^3 \pi^{ij}_{(3)} + \ep^5 \pi^{ij}_{(5)}  \rb  + O(\ep^6) \,.
\eeq
(See the complete equation~(17) in~\cite{Tichy:2002ec}.) 
Explicit expressions for $\pi^{ij}_{(3)}$ and $\pi^{ij}_{(5)}$ can be found in \cite{Jaranowski:1997ky};
The leading-PN-order term, $\pi^{ij}_{(3)}$, coincides with the standard Bowen-York linear-momentum
contribution for a nonspinning binary.

Near each particle, the spatial metric can be approximated by:
\beq
\ga^{\rm PN}_{ij}\sim \lb1+\fr{E_a}{2r_a} \rb^4 \de_{ij} + O(1/r_a^3) \,,
\eeq
which is just the Schwarzschild metric in isotropic coordinates.
 For $r_a\to 0$, the coordinate singularity is approached. This represents the
inner asymptotically flat end of the Schwarzschild metric in isotropic coordinates,
or the puncture representation of the Schwarzschild solution.
 This shows that if the metric form is kept as in equation~(\ref{eq:gapn}), then there
is a black hole centred on each particle.

The traceless-transverse part of the spatial three-metric, on the other hand,
can be constructed 
by imposing an outgoing wave condition.  We can rewrite the evolution equation
for $h^{\rm TT}_{ij}$ as:
\beq
h^{\rm TT}_{ij} = - \Box^{-1}_{\rm ret} \delta^{{\rm TT}\,kl}_{ij} 
\left[ \sum_{a=1}^N \, \frac{p_{ak} \, p_{al}}{m_a} \delta(x-x_a) 
+ \frac{1}{4} \phi^{(2)}_{,k} \phi^{(2)}_{,l} \right] \,,\label{eq:hTT_general}
\eeq
where $\de_{kl}^{{\rm TT}ij}$ is the TT-projection operator as defined in~\cite{Schaefer85}.
   Sch\"{a}fer~\cite{Schaefer85} suggested a ``near zone'' approximation
for $h^{\rm TT}_{ij}$, by splitting the retarded inverse d'Alembertian
in \eqref{eq:hTT_general} with an inverse Laplacian:
\bea
h^{\rm TT}_{ij} &=& - [ {\Delta^{-1}} + (\Box^{-1}_{\rm ret}  - \Delta^{-1} ) ] \delta^{{\rm TT}\,kl}_{ij} \left[ \cdots \right]
\nonumber \\
            &=& {h^{\rm TT\,(NZ)}_{i j}} + h^{\rm TT\,(remainder)}_{i j} + O(\ep^5) \,.
\eea
The explicit form of this near-zone (NZ) approximation was given by Jaranowski and
Sch\"{a}fer~\cite{Jaranowski:1997ky}.
A few years later, Kelly \etal~\cite{Kelly:2007uc}
completed the picture for nonspinning black holes by determining the ``remainder''
TT term, $h^{\rm TT\,(remainder)}_{i j}$ to 2PN order.
The structure of the remainder term divides into three segments,
according to time of evaluation:
\beq
h^{\rm TT\,(remainder)}_{i j} = h^{\rm TT\,(present)}_{i j} + h^{\rm TT\,(retarded)}_{i j} + h^{\rm TT\,(interval)}_{i j} \,.
\eeq
For each field point where $h^{\rm TT}_{i j}$ is to be evaluated,
the ``present'' term is evaluated using the particle positions and momenta
at $t=0$. The ``retarded'' term is evaluated using positions and momenta at
the retarded time of each source particle relative to the field point.
 The ``interval'' term is an integral over the particles' paths from
the retarded time to the present.

The present-time piece almost completely \emph{cancels} the near-zone
solution. The kinetic terms (i.e., those involving particle momenta)
cancel exactly, while potential terms (those involving particle-pair
separations) are strongly suppressed.
 The retarded-time piece reduces to the quadrupole solution for a nonspinning binary
as $r_{12}/r \rightarrow 0$ where $r$ denotes the field distance.
 The interval piece is too difficult to do in generality: we must integrate numerically.

The ADM-TT data introduced here has several attractive properties, which we briefly
restate here. The three-metric and extrinsic curvature expressions are easily found.
Unlike harmonic coordinates, ADM-TT has no logarithmic divergences. For a single BH,
the data reduces to the Schwarzschild metric in isotropic coordinates. Up to 1.5PN order,
the data coincides with the (unsolved) puncture approach -- conformally flat, with
Bowen-York extrinsic curvature. The trace of the extrinsic curvature vanishes up to 3PN
order. 
The Hamiltonian constraint decouples from the momentum constraints.

%%%%%%%%%%%%%%%%%%%%%%%%%%%%%%%%%%%%%%%%%%%%%%%%%%%%%%%%%%%%%%%%%%%%%%%%
\section{Encoding the Past History of the Binary}\label{sec:PNtraj}
%%%%%%%%%%%%%%%%%%%%%%%%%%%%%%%%%%%%%%%%%%%%%%%%%%%%%%%%%%%%%%%%%%%%%%%%

To evaluate this initial data for a given separation at $t = 0$, we must know not
only the particle momenta at $t=0$, but also the position and momentum of each
particle over the past history of the binary, at least from the retarded time
relative to the most distant grid point in the numerical domain. That is, the larger
the numerical domain, the further back in time we must look for the information
needed to fill in the $h_{ij}^{\rm TT}$ fields at these spatially distant points.

To supply this position and momentum information, we evolve the Hamiltonian equations
of motion, using a standard Taylor PN Hamiltonian 
and flux function evaluated at 3PN and 3.5PN order, respectively. 
The set of equations which we use in this paper 
has been given in~\cite{Buonanno:2005xu}. 
This evolution is started at a suitably large separation
($r_{12} \sim 40M$ is easily sufficient for current purposes), and the separation
and orbital phase are saved for later interpolation. We then apply a shift in time
and orbital frequency to match the initial conditions of the fully numerical evolution.

This shifted trajectory data is used directly to evaluate the present-time fields. We
also use it to evaluate the two retarded times (one for each particle) at each field point
by a nonlinear Newton-search algorithm. Then we can evaluate the retarded-time fields,
and numerically integrate the interval terms.

%%%%%%%%%%%%%%%%%%%%%%%%%%%%%%%%%%%%%%%%%%%%%%%%%%%%%%%%%%%%%%%%%%%%%%%%
\section{Hybrid Initial Data}\label{sec:HID}
%%%%%%%%%%%%%%%%%%%%%%%%%%%%%%%%%%%%%%%%%%%%%%%%%%%%%%%%%%%%%%%%%%%%%%%%

 The similarities of the wavy ADM-TT leading-order terms and the Bowen-York
puncture data may be the key to further constraining the initial data.
We use the rationale that the largest Hamiltonian violation contributions are due to the
PN conformal factor, 
$\psi_{\rm PN}$, since $h^{\rm TT}_{ij}$ is relatively much smaller for field points 
closer to the punctures.
We propose then a hybrid initial data prescription in which we first solve
the Hamiltonian constraint for the traditional Bowen-York puncture initial data
and then rescale and superpose the higher-order PN terms to this solution.
More specifically, the Bowen-York extrinsic curvature is used as usual to source
the Hamiltonian constraint for a conformally flat spatial metric:
\beq
\De \psi_{\rm BY} +\frac{1}{8} K^{ij}_{\rm BY} K_{ij}^{\rm BY}\psi_{\rm BY}^{-7} = 0 .
\eeq
The solution follows then from the puncture trick and is given by
\beq
\psi_{\rm BY} = 1 + \frac{1}{2} ( \frac{m_1}{r_1} + \frac{m_2}{r_2}) + u,
\eeq
where $u$, the regular part of the solution, is determined by the numerical solution
of the resulting Poisson-like elliptic equation. The functional similarity of this solution 
to the one provided by the PN expansion, equation~\eqref{eq:psiPN}, is quite striking.
The singular part, or the Brill-Lindquist part~\cite{Brill:1963yv} if you wish, has the
same functional form in each case, differing only in how the bare masses of particles
and black holes are treated, and the missing regular contribution $u$ in the PN 
conformal factor.  In this new approach the ADM canonical quantities are then rescaled 
by the Bowen-York solution, $\psi_{\rm BY}$, instead of the PN conformal factor
$\psi_{\rm PN}$:
\bea
\ga_{ij} &=& \psi_{\rm BY}^4 \de_{ij} + h^{\rm TT}_{ij}  \,,  \\
\pi^{ij} &=& \fr{1}{\psi_{\rm BY}^4} \lb \ep^3 \pi^{ij}_{(3)} + \ep^5 \pi^{ij}_{(5)}  \rb \,,
\eea
where we note that $\pi^{ij}_{(3)}$ corresponds exactly to the spinless Bowen-York 
contribution. The transverse-traceless part of the metric, $h^{\rm TT}_{ij}$, 
and the higher-order PN contribution to the conjugate momentum,
$\pi^{ij}_{(5)}$, are then superposed with this puncture data.

%%%%%%%%%%%%%%%%%%%%%%%%%%%%%%%%%%%
\subsection{Constraint violations}
%%%%%%%%%%%%%%%%%%%%%%%%%%%%%%%%%%%

\begin{figure}
 \begin{center}                                                                 
   \includegraphics[width=7cm]{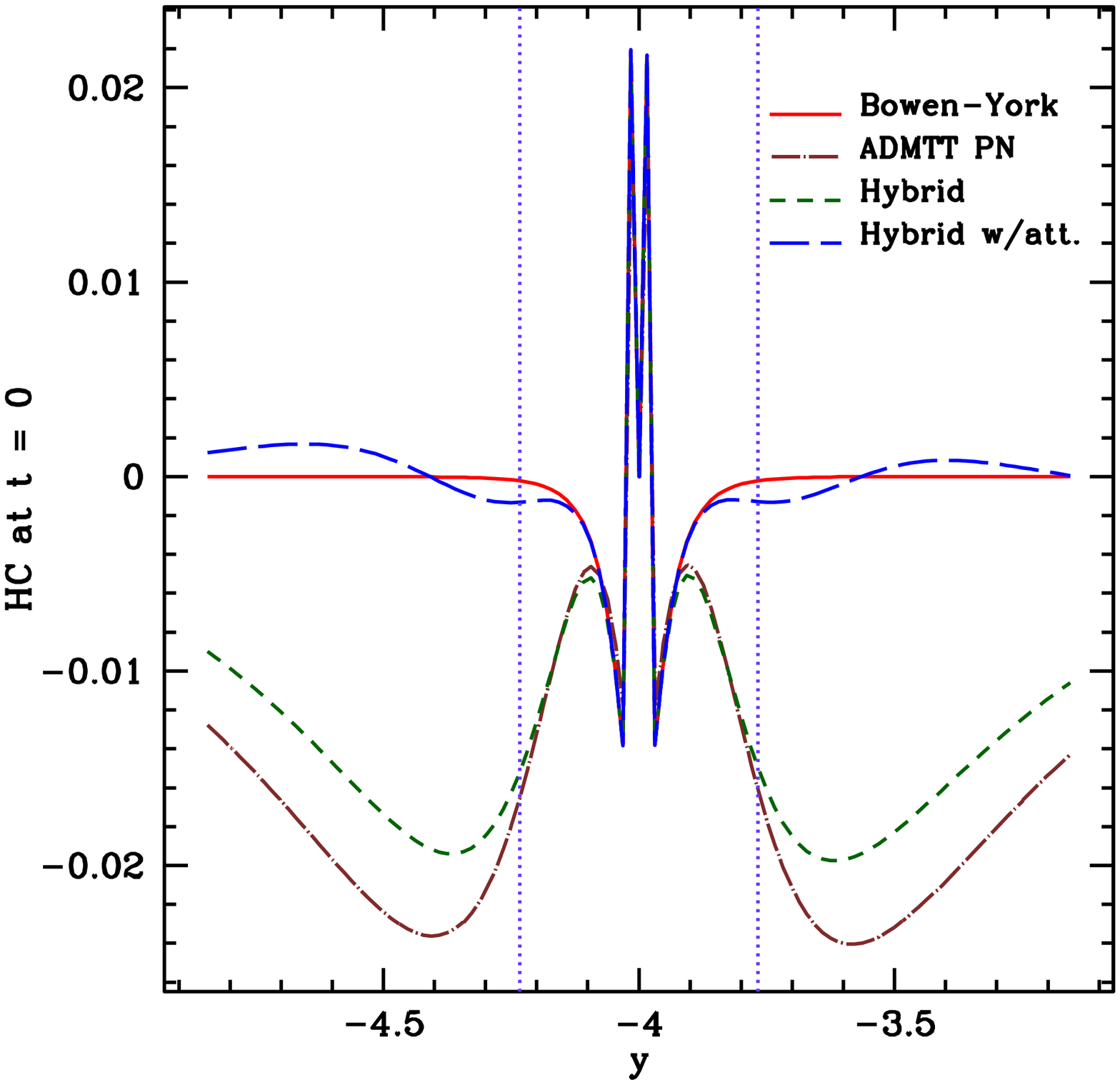}    
   \includegraphics[width=7cm]{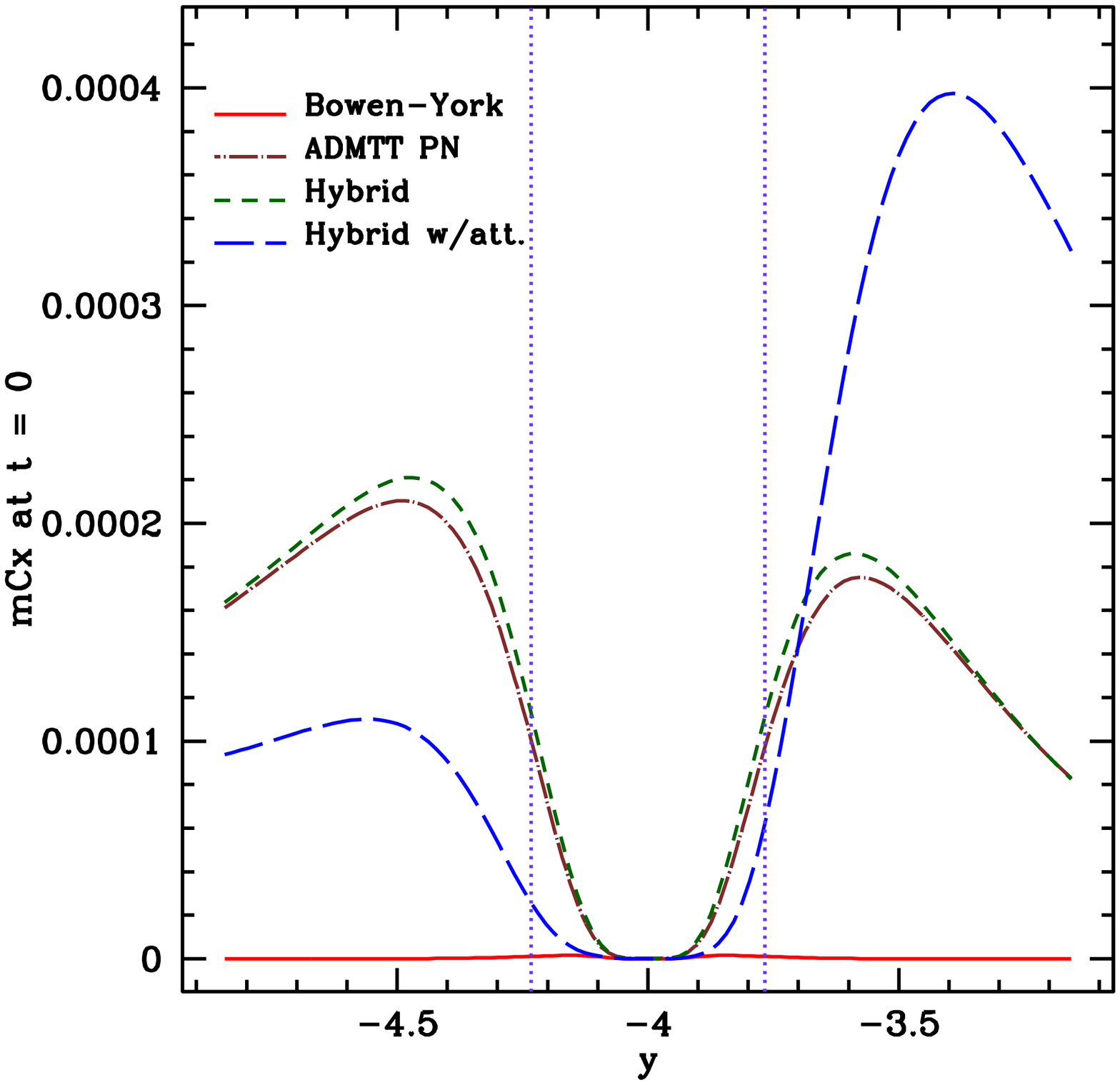}    
 \end{center}
 \caption{Constraint violations. These figures compare the violation of the constraint 
equations around a puncture located in the $y$-axis at $y$=$-4M$.  The (purple) dotted 
vertical lines indicate the location of the apparent horizon. The (red) solid line represents
the violation for the traditional Bowen-York data.  The long-dash-dotted (brown) lines
describes the violation for the ADM-TT PN data, while the (green) dashed and
the (blue) long-dashed lines correspond to the Hybrid and Hybrid with attenuation function,
respectively.  We keep the line styles consistently the same in all figures hereafter 
to facilitate comparison.  The plot on the left panel represents the Hamiltonian 
constraint violation and the one on the right represents the momentum constraint violation.}
 \label{fig:constraint_comps}
\end{figure}

Looking into the violation of the constraint equations, Fig.~\ref{fig:constraint_comps}, 
it becomes clear that the PN metric becomes inaccurate close to the punctures, 
at least when compared to the traditional Bowen-York data.  The long-dash-dotted brown curve
there represents the constraint violation for the PN metric in the ADM-TT 
gauge, labeled in the figure and in all plots hereafter as ``ADMTT PN''. The purple
vertical dotted lines indicate the apparent horizon location around one of the 
punctures; here the one located on the $y$-axis at $y$=$-4M$. Focusing our attention
on the plot on the left panel, the Hamiltonian constraint violation, we can observe 
a negative violation outside of the horizon for the PN data.  This can
be interpreted as, and has the effect of, an unphysical negative mass gravitating around
each black hole.  Unfortunately, the hybrid approach just introduced in the last section
did not reduce this violation considerably, as the green dashed line in the plot shows.
We decided then to investigate a way to improve the data approximation closer to the
apparent horizons and that led to the second crucial element of the approach 
proposed here: the use of attenuation functions.

%%%%%%%%%%%%%%%%%%%%%%%%%%%%%%%%%%%%%%%%%%%%%%%%%%%%%%%%%%%%%%%%%%%%%%%%
\subsection{Attenuation function}\label{sec:atte}
%%%%%%%%%%%%%%%%%%%%%%%%%%%%%%%%%%%%%%%%%%%%%%%%%%%%%%%%%%%%%%%%%%%%%%%%

Since the PN approximation of the metric and extrinsic curvature
is inaccurate for the inner or internal zone, i.e., the region around each 
black hole in binary systems where $r_1 \ll r_{12}$ and $r_2 \ll r_{12}$, 
we need to consider a different approach 
to construct the initial data around that region.
Johnson-McDaniel, Yunes, Tichy and Owen~\cite{JohnsonMcDaniel:2009dq} 
introduced the black hole perturbation calculation by 
Detweiler~\cite{Detweiler:2005kq} to be used as an approximation 
for the inner zone.  

\begin{figure}
 \begin{center}
    \includegraphics[width=13cm]{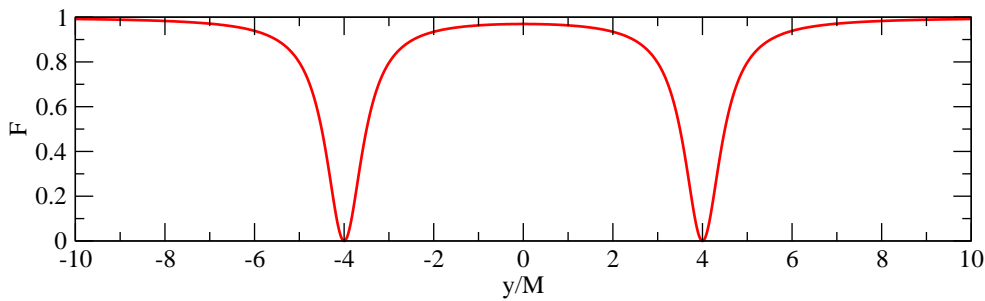}
 \end{center}
   \caption{The attenuation function along the $y$ axis 
   for a binary with $m_1=m_2=M/2$ at $y_1=-y_2=4M$,  
   and $a_t=1$.}
 \label{fig:atte}
\end{figure}

Here, we treat the inner-zone metric by a simpler approach. 
We discard the PN contribution around each black hole by applying 
an attenuation function defined as follows 
(see Fig.~\ref{fig:atte} for its functional form along an interval on the $y$-axis): 
\begin{eqnarray}
{\cal F} &= \frac{1}{4}\, 
\left[\tanh \left( \ln \left(\frac{r_1}{a_t \,m_1} \right) \right)+1 \right]
\left[\tanh \left( \ln \left(\frac{r_2}{a_t \,m_2} \right) \right)+1 \right]
\nonumber \\
&= \left( 1 + \frac{{a_t}^2 \,{m_1}^2}{{r_1}^2} \right)^{-1}
\left( 1 + \frac{{a_t}^2 \,{m_2}^2}{{r_2}^2} \right)^{-1}
\,, 
\end{eqnarray}
where $a_t$ is a constant parameter\footnote{ 
Note that when we apply this attenuation function to $h^{\rm TT}_{ij}$, 
the correction at large distances is at higher PN order 
than we treat in this paper. 
At large distances, we a Taylor expansion with respect to 
$m_i/r_i$ (i.e., a PN expansion) yields 
\begin{eqnarray}
{\cal F} h^{\rm TT}_{ij} = h^{\rm TT}_{ij} \,
\left(1 - \frac{{a_t}^2 \,{m_1}^2}{{r_1}^2} - \frac{{a_t}^2 \,{m_2}^2}{{r_2}^2} 
+ \cdots \right) \,.
\end{eqnarray}
Therefore, the corrections due to the attenuation function 
are at 2PN higher order than the leading order of $h^{\rm TT}_{ij}$.}.
Our hybrid initial data approach with the use of an 
attenuation function then becomes:
\bea
\ga_{ij} &=& \psi_{\rm BY}^4 \de_{ij} + {\cal F} h^{\rm TT}_{ij}  \,,  \\
\pi^{ij} &=& \fr{1}{\psi_{\rm BY}^4} \lb \ep^3 \pi^{ij}_{(3)} + {\cal F} \ep^5 \pi^{ij}_{(5)}  \rb \,.
\eea
The effect of this procedure is a considerable reduction of the Hamiltonian
constraint violation, as the blue long-dashed curve indicates on the left panel
of Fig.~\ref{fig:constraint_comps}.

While the Hamiltonian constraint violation is improved, and we are going to 
discuss its effect on the data evolution later on, the momentum constraint violation
seems to be unaffected by these two key ingredients of our procedure. Both the 
Hybrid and the Hybrid data with attenuation seem to affect the momentum constraint
residuals negligibly when compared to the PN data. We are also going 
to discuss later how we think this may be affecting the data evolution.
But before discussing any further properties of this data, let us say a few words
on how the numerical experiments were prepared and performed. This is the topic
of the next section.

%%%%%%%%%%%%%%%%%%%%%%%%%%%%%%%%%%%%%%%%%%%%%%%%%%%%%%%%%%%%%%%%%%%%%%%%
\section{Numerical Evolution}\label{sec:num}
%%%%%%%%%%%%%%%%%%%%%%%%%%%%%%%%%%%%%%%%%%%%%%%%%%%%%%%%%%%%%%%%%%%%%%%%

The black-hole binary studied in these experiments had a mass ratio
of $q=1$, was nonspinning, with punctures initially located on the $y$-axis at $\pm 4M$.
In this framework we use the puncture
approach~\cite{Brandt:1997tf} along 
with the {\sc TwoPunctures} thorn \cite{Ansorg:2004ds}
to calculate the Bowen-York initial data. We evolved these
black-hole-binary data-sets using the {\sc LazEv}~\cite{Zlochower:2005bj} 
implementation of the moving puncture
formalism~\cite{Campanelli:2005dd,Baker:2005vv} with the conformal
factor $W=\sqrt{\chi}=\exp(-2\phi)$ suggested by~\cite{Marronetti:2007wz}
as a dynamical variable.
For the runs presented here
we use centred, eighth-order finite differencing in
space~\cite{Lousto:2007rj} and an RK4 time integrator (note that we do
not upwind the advection terms).

We obtain accurate, convergent waveforms and horizon parameters by
evolving this system in conjunction with a modified $1+\log$ lapse and a
modified Gamma-driver shift
condition~\cite{Alcubierre:2002kk,Campanelli:2005dd}, and an initial lapse
$\alpha(t=0) = 2/(1+\psi_{\rm BL}^{4})$.  The lapse $\alpha$ 
and shift $\beta^a$ are evolved with
\begin{eqnarray}
(\partial_t - \beta^i \partial_i) \,\alpha = - 2 \,\alpha \,K \,, 
\nonumber \\
 \partial_t \,\beta^a = \frac{3}{4} \,\tilde \Gamma^a - \eta \,\beta^a \,,
 \label{eq:gauge}
\end{eqnarray}
where $\eta$ is constant.

We use the Cactus/Einstein Toolkit code~\cite{cactus_web,einsteintoolkit_web} 
to provide the parallel infrastructure and the Carpet~\cite{Schnetter:2003rb} 
mesh refinement driver to provide a `moving boxes' style mesh refinement. 
In this approach refined grids of fixed size are arranged about the coordinate centers
of both holes.  The Carpet code then moves these fine grids about the
computational domain by following the trajectories of the two black
holes.

We use the {\sc AHFinderDirect} thorn~\cite{Thornburg:2003sf} to locate
apparent horizons.  We measure the magnitude of the horizon spin $S$ using
the Isolated Horizon algorithm detailed in~\cite{Dreyer:2002mx}.
Once we have the
horizon spin, we can calculate the horizon mass $\MAH$ via the Christodoulou
formula \cite{Christodoulou:1970wf}:
\beq
\MAH = \sqrt{\Mirr^2 + \frac{S^2}{4 \,\Mirr^2}} \,,
\eeq
where $\Mirr = \sqrt{A/(16 \pi)}$ is the \emph{irreducible mass} and $A$
is the surface area of the horizon.

\begin{figure}
 \begin{center}                                                                 
   \includegraphics[angle=-90,width=8cm]{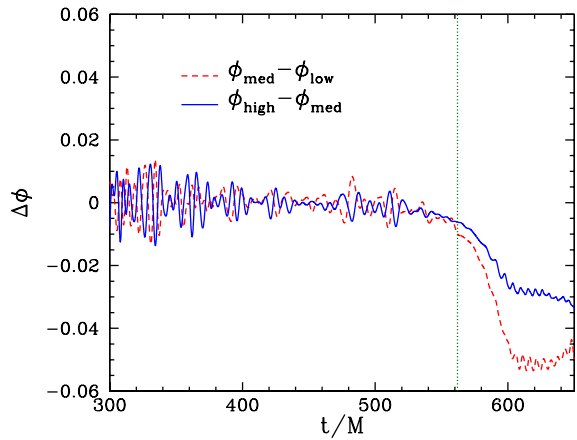}    
 \end{center}
 \caption{Phase accuracy. This figure shows the difference in phase for the $(2,2)$
 mode of the radiative Weyl scalar $\psi_4$ at different resolutions. We compare the
 difference of the medium and low resolutions 
 with that of the high and medium ones. We see that 
 the phase error is below the threshold of $0.05$ at $M\Om=0.2$ 
 (vertical dotted green line) suggested by the NRAR collaboration~\cite{nrar_web}.
}
 \label{fig:accuracy}
\end{figure}

For the runs in this paper, we used a grid hierarchy spanning nine levels of refinement. 
We ran three different sets of grid hierarchy resolutions in order to determine the
waveform accuracy, shown in Fig.~\ref{fig:accuracy}. The low-, medium- and high-resolution hierarchies have
mesh spacings of $h_c = \{5M,4M,3.2M \}$ for the coarsest grid and 
of $h_f = \{M/51.2,M/64,M/80 \}$ for the finest ones, respectively.
The mesh spacings in the wave-extraction region are $h_{\rm ext} = \{1.25M,1M,0.8M \}$.
We apply reflection-boundary symmetry at the $z=0$ plane and $\pi$ symmetry (that is, particle-exchange symmetry)
at the $x=0$ plane. The outer boundary is located at $400M$. 

The results comparing the different initial data sets, presented in the following sections,
used the medium grid hierarchy only, i.e.,
the finest mesh spacing for this hierarchy is $h_f=M/64$,
 resulting in approximately 
$30$ grid points across the initial apparent horizon.

%%%%%%%%%%%%%%%%%%%%%%%%%%%%%%%%%%%
\subsection{Apparent Horizon Mass}
%%%%%%%%%%%%%%%%%%%%%%%%%%%%%%%%%%%

\begin{figure}
 \begin{center}                                                                 
   \includegraphics[angle=-90,width=12cm]{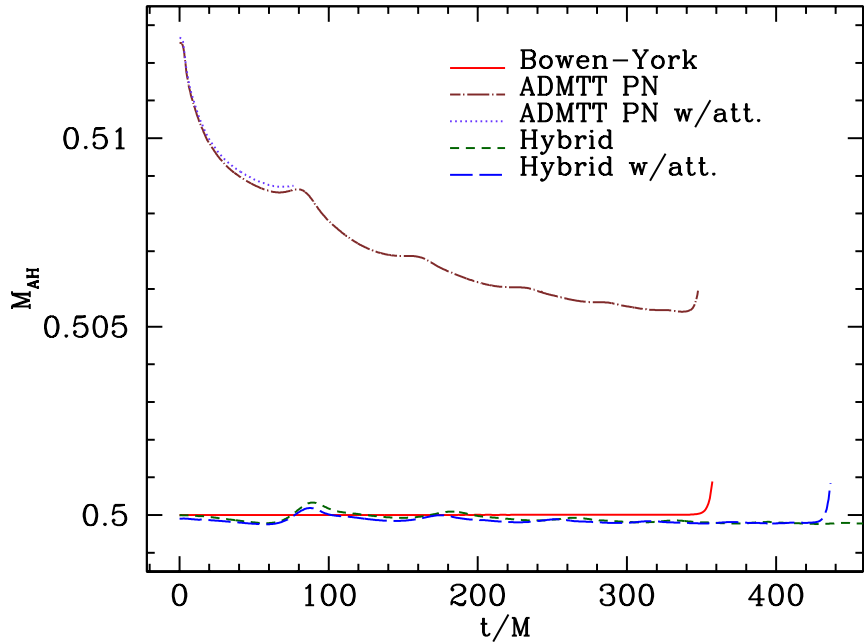}    
 \end{center}
 \caption{The behaviour of each hole's apparent horizon mass $\MAH$
 for Bowen-York, ADM-TT PN, ADM-TT PN with attenuation [(purple) dotted line], 
 Hybrid and Hybrid with attenuation initial data. 
 The mass ``leak'' was reduced considerably with the hybrid initial data.}
 \label{fig:ahm}
\end{figure}

The first result from the evolution of the data introduced in the previous 
section that we would like to discuss concerns the apparent horizon mass.
It was reported earlier~\cite{Kelly:2009js} for the PN data that the mass was not
being conserved with time -- some sort of mass ``leak'' was plaguing 
the data evolution. We confirm this problem in Fig.~\ref{fig:ahm}, as we can 
observe the long time it takes for the PN irreducible mass to relax towards
a constant value. 

The first step we took when trying to tackle this issue was to introduce 
the attenuation function for the PN data. As the purple dotted 
curve in the figure reveals, this approach alone was not successful. Only
when we introduced the hybrid approach to the binary black-hole initial data,
did we see an improvement in this unwanted feature.
Now the mass conservation seems much more accurate, left only the presence
of some bumps in the curve.

We believe that the presence of these bumps in the data are related to 
unphysical negative mass, resulting from the initial Hamiltonian
constraint violation. Since these bumps appear  
approximately every half orbital period, or roughly when the punctures
cross the $y$-axis, it may be reasonable to assume that the initial
Hamiltonian violations hangs in the domain around their initial locations
or in the holes' ``wakes'', occasionally being absorbed by the passing hole.  
The attenuation function seems to reduce this effect slightly, however only 
when associated with the hybrid data.

%%%%%%%%%%%%%%%%%%%%%%%%%%%%%%%%%%%
\subsection{Eccentricity}
%%%%%%%%%%%%%%%%%%%%%%%%%%%%%%%%%%%

Several methods have been proposed to estimate the eccentricity of numerically evolved
binary inspirals; for recent examples, see \cite{Buonanno:2006ui,Mroue:2010re,Tichy:2010qa}.
We estimate the eccentricity of all four different data sets by first $\chi^2$-fitting
a polynomial function, $r_{\rm fit}(t)$, to the binary coordinate separation, $r_{\rm NR}(t)$, 
as a function of time. We subtract then the numerical data from the fitted function and 
normalize it by the same fitted function:
\beq
e_r(t)=\frac{r_{\rm NR}(t)-r_{\rm fit}(t)}{r_{\rm fit}(t)} \simeq e_r \cos(\Omega_r t + \phi_0) \,.
\label{eq:e_r}
\eeq
The interpolating polynomial must be monotonically decreasing with no extrema in the fitting
interval. We use the highest possible polynomial order satisfying these constraints.
The assumption behind this procedure is a secular quasi-adiabatic shrinking of the orbital
separation. The amplitude of the sinusoidal part extracted out of this orbital decay 
provides then an estimate of the orbital eccentricity. 

\begin{figure}
 \begin{center}
    \includegraphics[angle=-90,width=7.7cm]{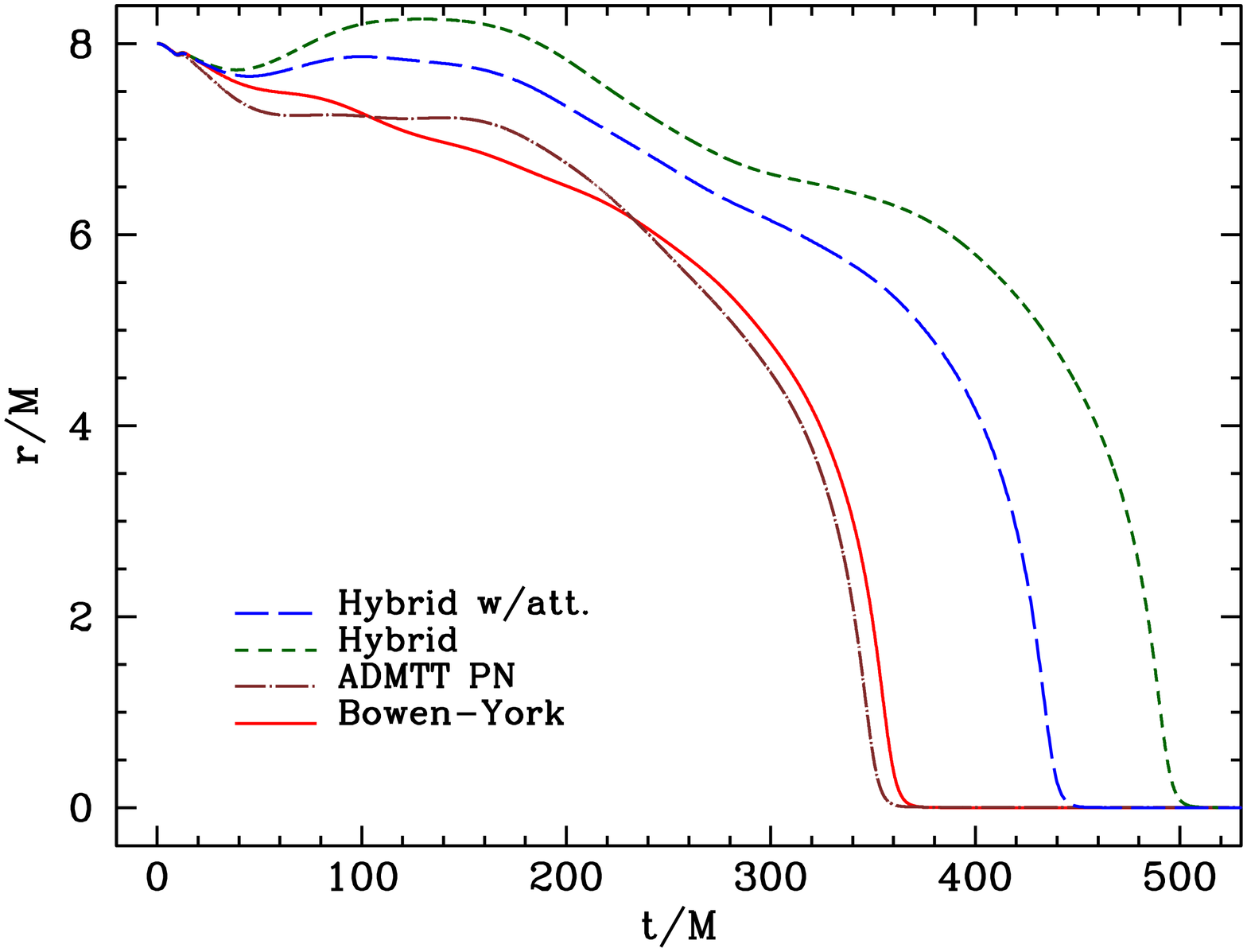}
    \includegraphics[angle=-90,width=7.7cm]{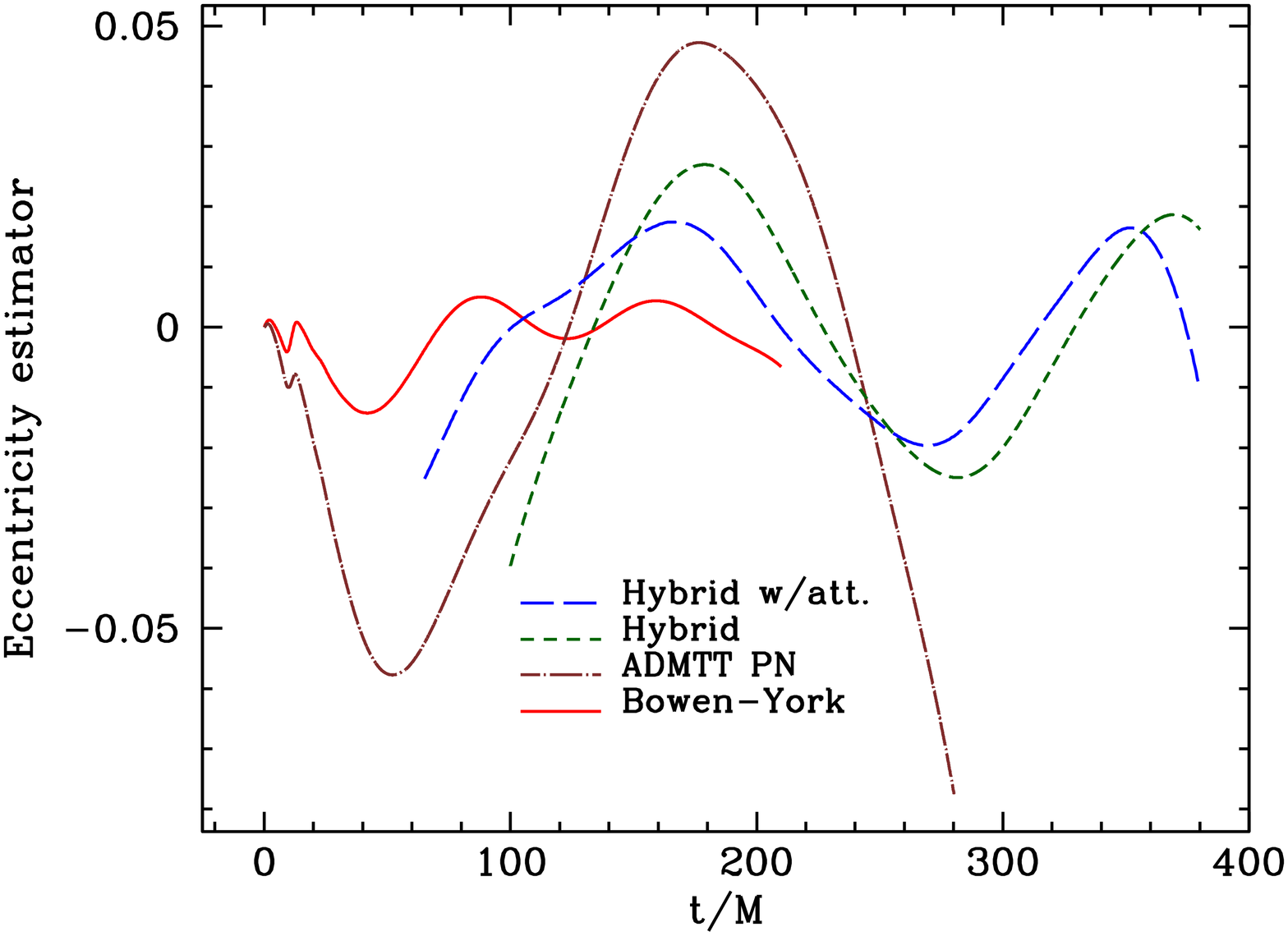}
 \end{center}
 \caption{Coordinate separation (left panel) and 
 its eccentricity estimator in equation (\ref{eq:e_r})
 (right panel) versus time for Bowen-York, ADM-TT PN, Hybrid and 
 Hybrid with attenuation initial data.}
\label{fig:separation}
\end{figure}

Specifically for these runs, where the initial separation is
$r_0=8M$, we fit a second-order polynomial curve to the numerical
data. We were careful to choose a time to stop fitting
much earlier than the coalescence time.  Usually we would eliminate
the initial gauge effects from the data, and start fitting somewhere
around $t=40M$ or later, as was the case for both instances of Hybrid data 
presented in Fig.~\ref{fig:separation}. However this procedure turned out to be 
very sensitive to the initial fit separation for the Bowen-York and PN
data. We decided then to include this initial transient
in order to establish only an upper bound estimate for the eccentricity.
We believe that this is not necessary for binaries initially farther apart. 

Amplitude readings from the eccentricity estimator plot as well as from a bare inspection
of the coordinate separation as a function of time, Fig.~\ref{fig:separation}, 
indicate that the Bowen-York data has an eccentricity of the order of $10^{-2}$. 
Clearly the PN eccentricity is around $5$ times this value.
The Hybrid data alone, on the other hand, did not have a considerable impact on the 
eccentricity, possibly due to the fact that the momentum constraint violations 
were not improved as well. We could minimize a bit the high eccentricity effects 
by introducing the attenuation function, but this problem remains open.

%%%%%%%%%%%%%%%%%%%%%%%%%%%%%%%%%%%
\subsection{Waveforms}
%%%%%%%%%%%%%%%%%%%%%%%%%%%%%%%%%%%

\begin{figure}
 \begin{center}
    \includegraphics[width=7cm]{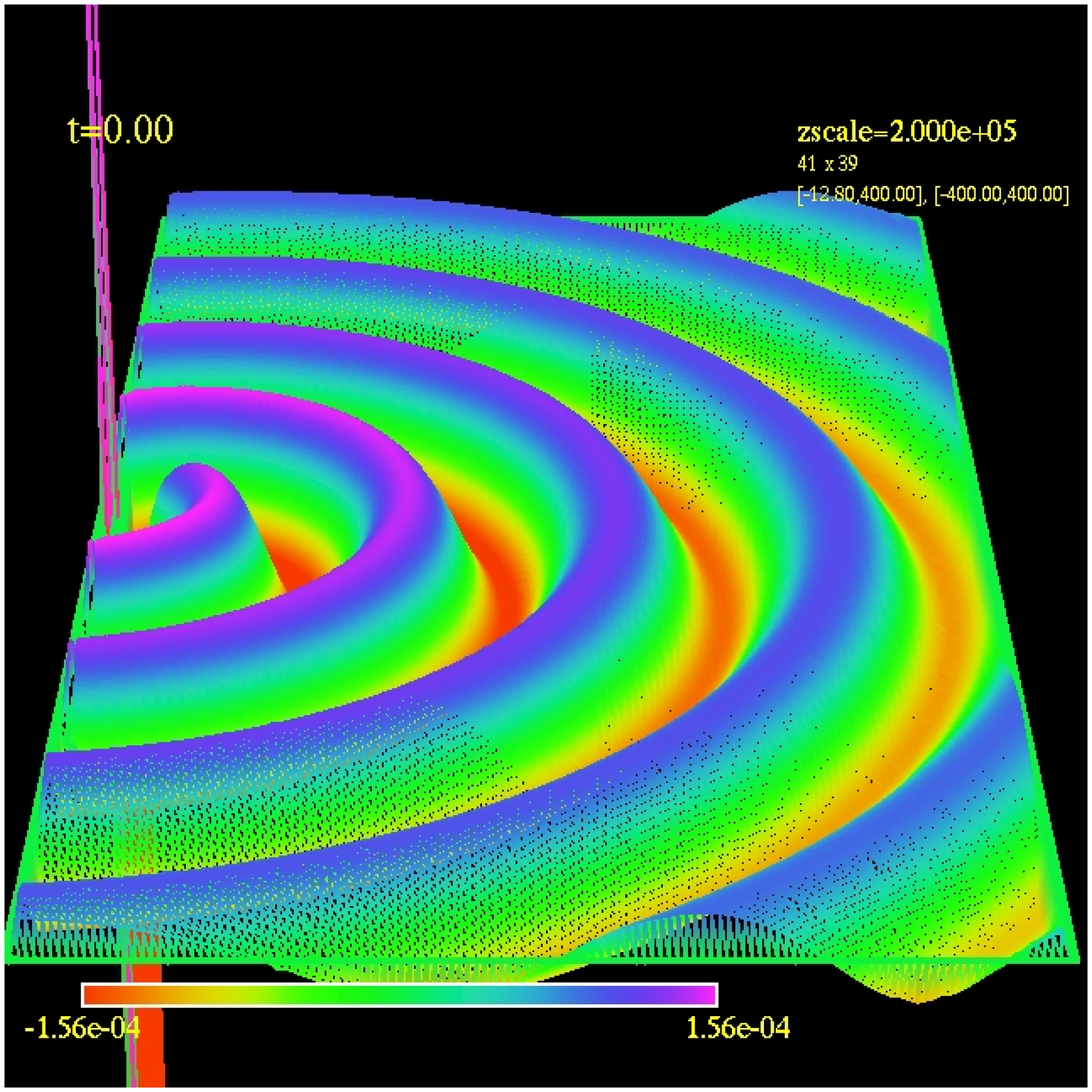}
    \includegraphics[width=7cm]{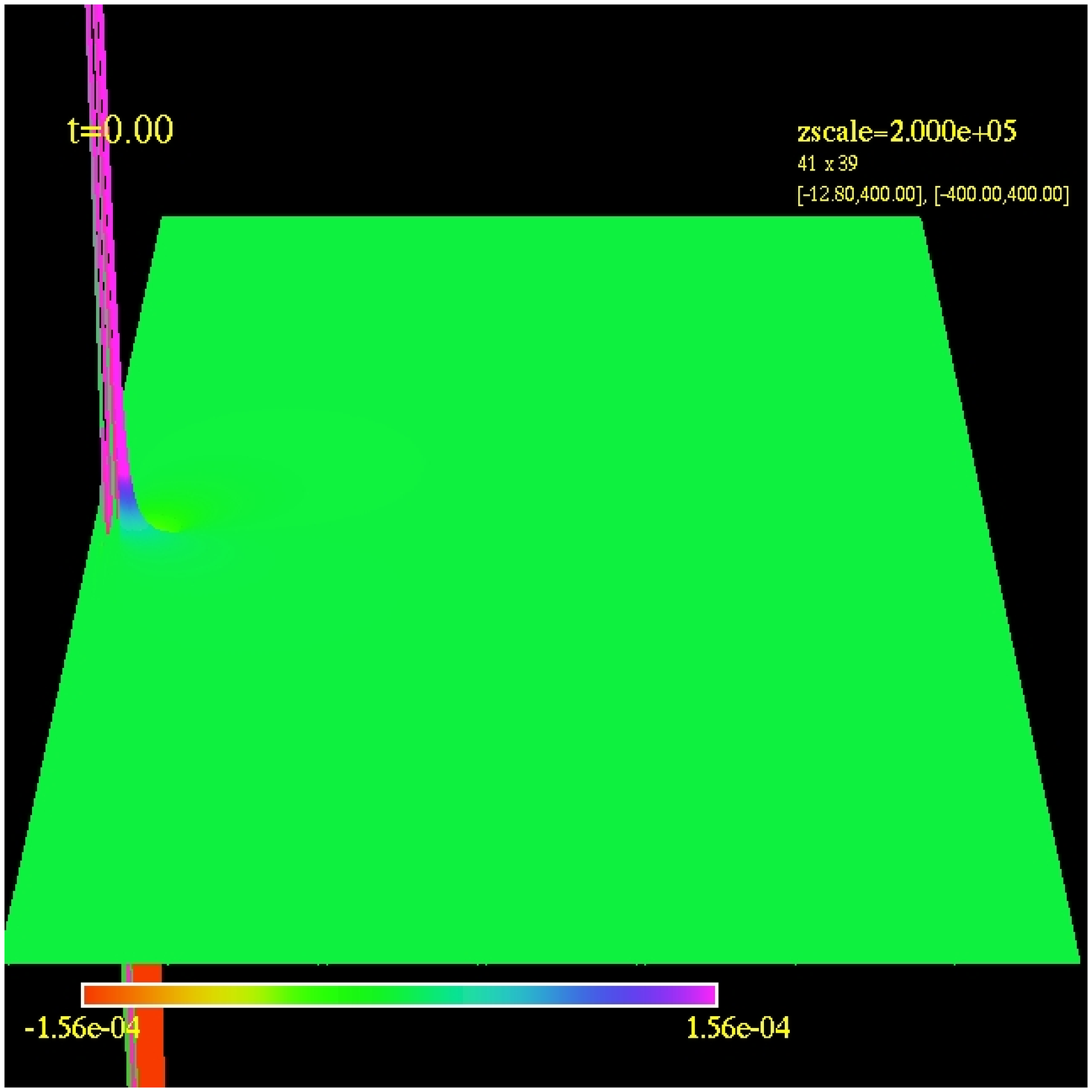}
 \end{center}
   \caption{Comparison between the wavy pattern at $t=0$ present in
   $R_{\rm areal} \,{\mathcal Re}\, (\psi_4)$ for the hybrid initial data (left) 
   and the traditional Bowen-York data (right).}
\label{fig:wavy}
\end{figure}

We illustrate in Fig.~\ref{fig:wavy} the presence of the realistic
wave content in the hybrid initial data.  The figure shows the 
$z=0$ slice of the data for the real part of the Weyl scalar $\psi_4$ 
multiplied by the areal radius $R_{\rm areal}$.
We show the traditional Bowen-York data for comparison in the right
panel of the same figure.

\begin{figure}
 \begin{center}
    \includegraphics[angle=-90,width=7.7cm]{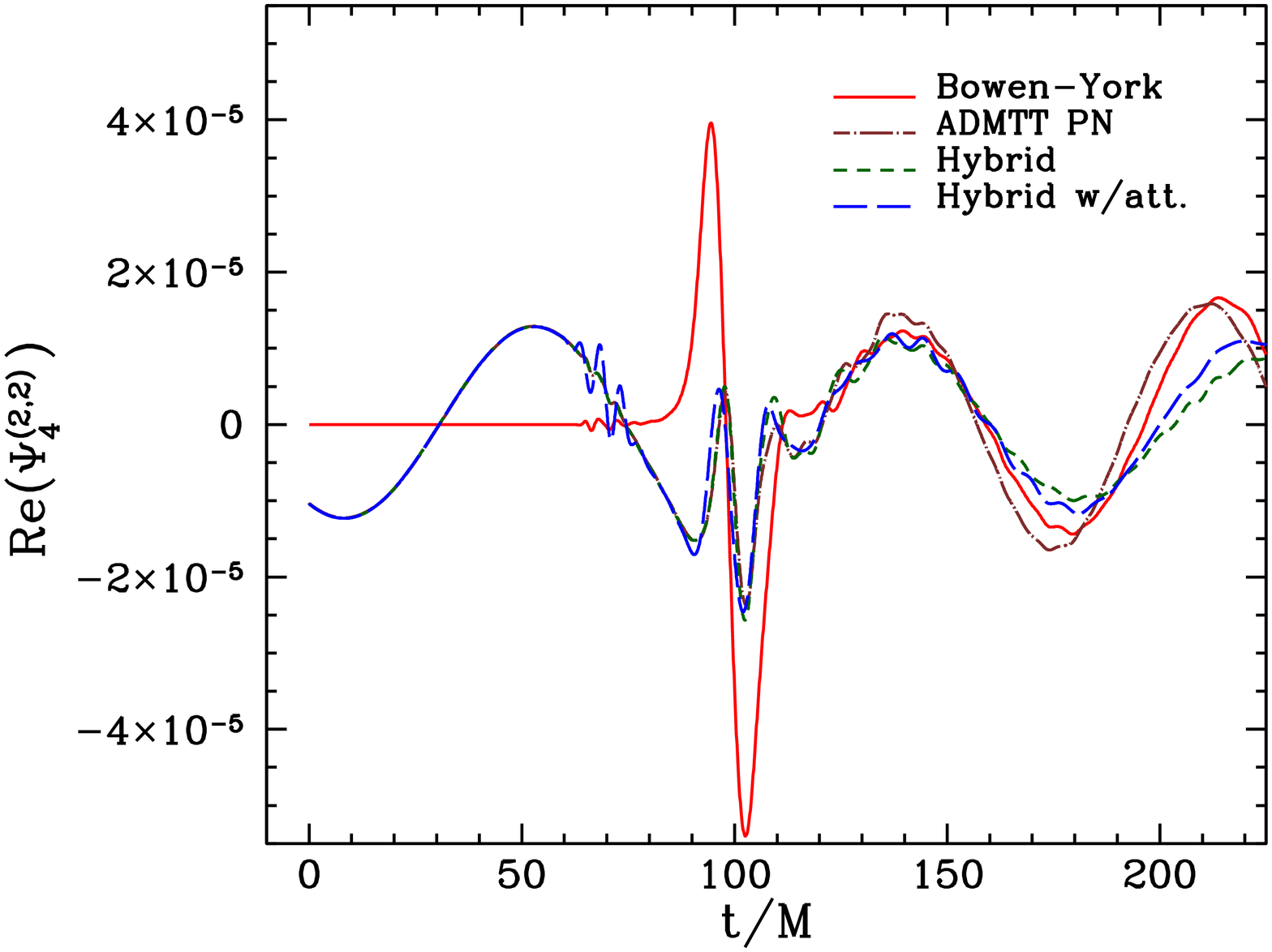}
    \includegraphics[angle=-90,width=7.7cm]{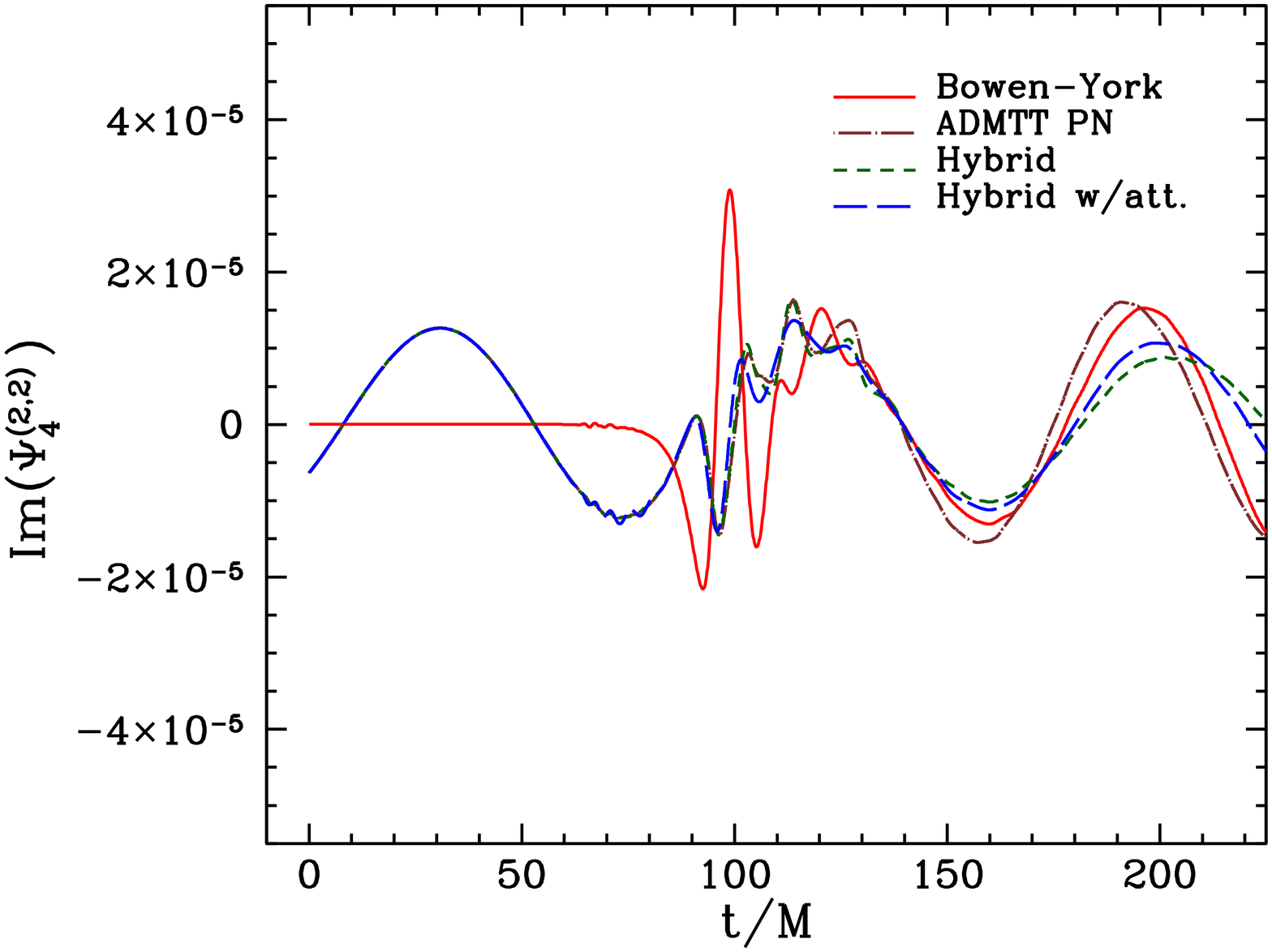}
 \end{center}
 \caption{Real and imaginary parts of the $(2,2)$ mode of the Weyl scalar $\psi_4$  
 resulting from the evolution of Bowen-York, ADM-TT PN, Hybrid and 
 Hybrid with attenuation initial data.  The extraction radius is $\Rext=90M$.}
\label{fig:ReIm}
\end{figure}

\begin{figure}
 \begin{center}
    \includegraphics[angle=-90,width=7.7cm]{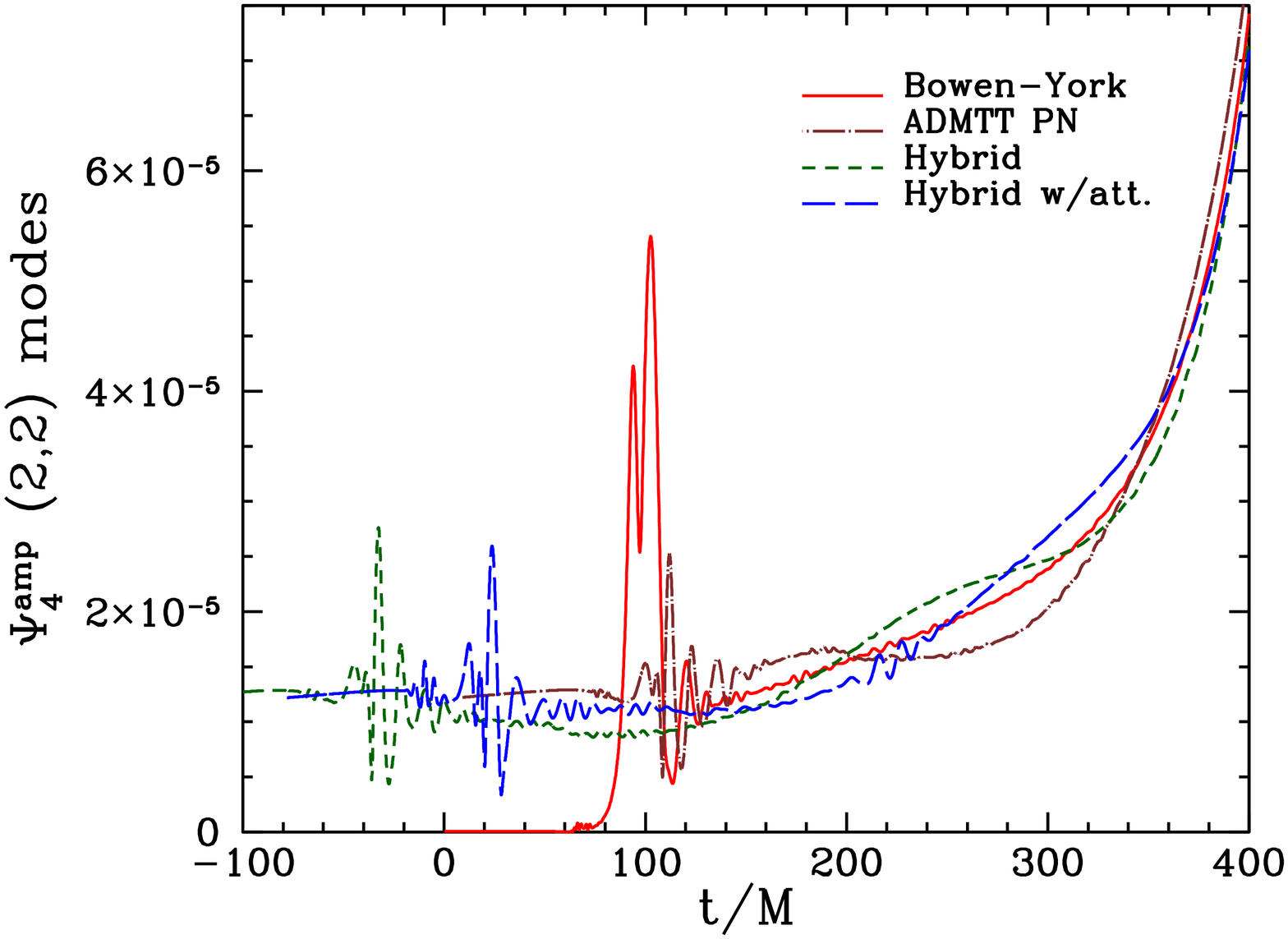}
    \includegraphics[angle=-90,width=7.7cm]{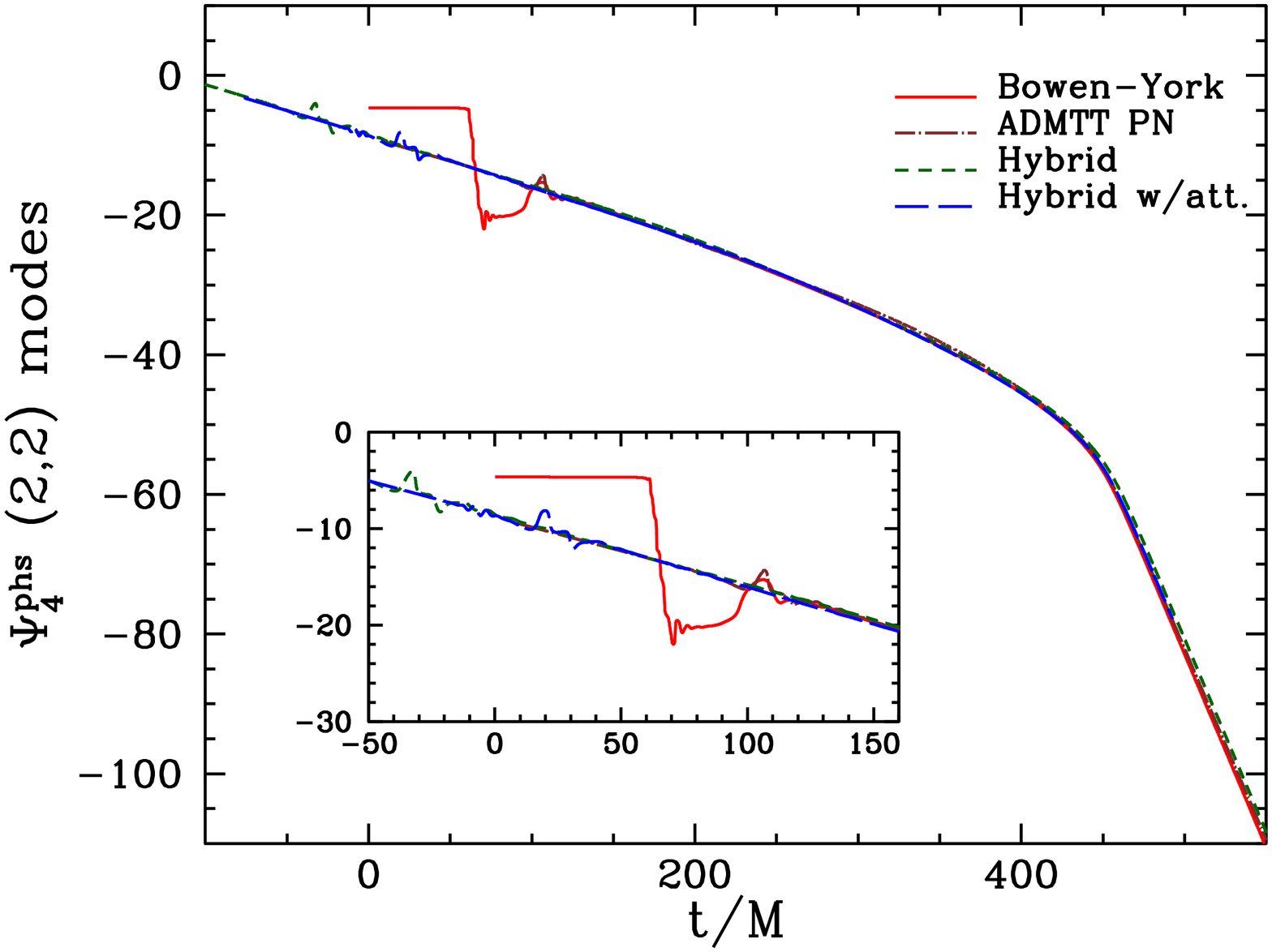}
 \end{center}
 \caption{Comparison between the $\psi_4$ $(2,2)$ mode amplitude (left) and 
 phase (right) for the Bowen-York, ADM-TT PN, Hybrid  and Hybrid 
 with attenuation initial data.  The extraction radius is $\Rext = 90M$.}
\label{fig:phs22}
\end{figure}

\begin{figure}
 \begin{center}
    \includegraphics[angle=-90,width=7.7cm]{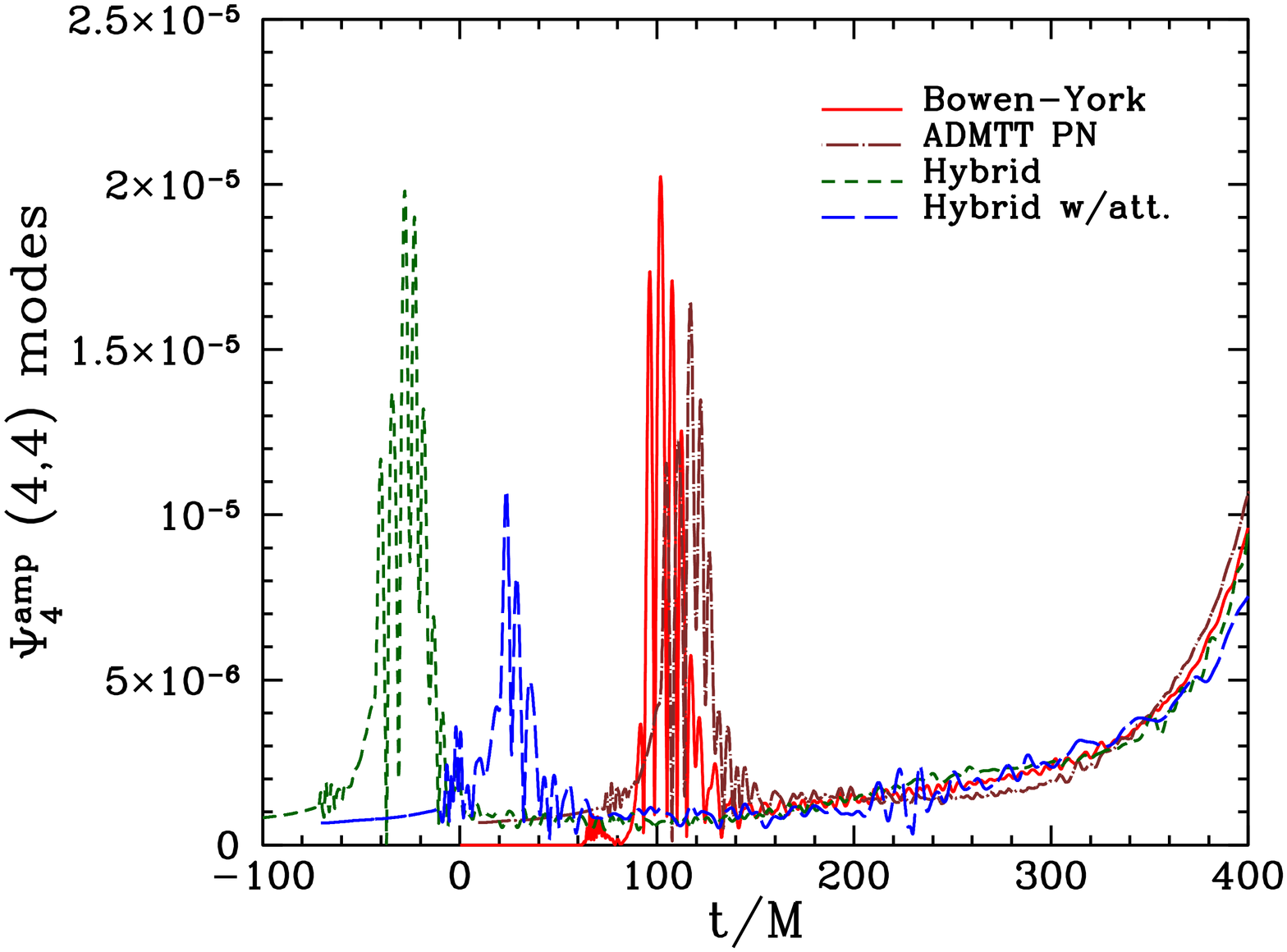}
    \includegraphics[angle=-90,width=7.7cm]{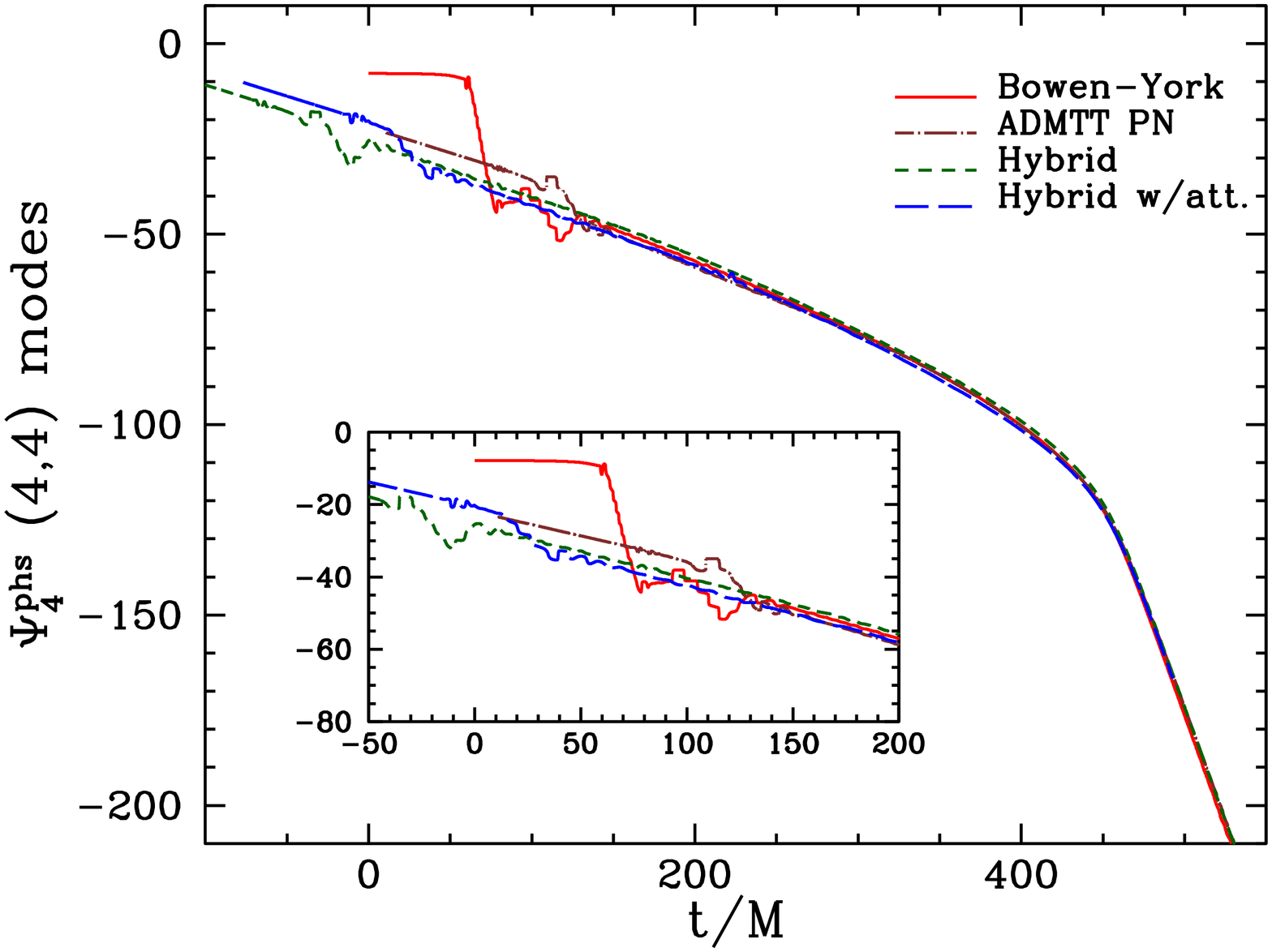}
 \end{center}
 \caption{Comparison between the $\psi_4$ $(4,4)$ mode amplitude (left) and 
phase (right) for the Bowen-York, ADM-TT PN, Hybrid  and Hybrid 
with attenuation initial data.  The extraction radius is $\Rext = 90M$.}
\label{fig:phs44}
\end{figure}

Like the wavy ADM-TT PN data, the hybrid data
significantly reduces the amplitude of the spurious radiation
when compared to the traditional Bowen-York data as is demonstrated in 
Fig.~\ref{fig:ReIm} for the real and imaginary parts of the $(2,2)$ mode 
of the Weyl scalar $\psi_4$. In Fig.~\ref{fig:phs22} we compare the amplitude (left panel) 
and phase (right panel) of this mode. 
We can observe a clear reduction in the amplitude and a dramatic improvement 
in the phase for all cases compared to Bowen-York data. Note that there is 
a decrease in amplitude for the Hybrid data, suggesting the binary is cast 
into a larger orbital separation. The use of attenuation function greatly 
reduces this effect, however.  For the $(4,4)$ mode we only achieve a visible 
reduction of the junk radiation in the amplitude with the Hybrid with attenuation data,
Fig.~\ref{fig:phs44}, which suggests that we need to improve the analytical 
approximation of the ADM canonical quantities in the inner zone if we 
want to accurately describe black-hole tidal effects.  The disturbances in 
phase still seem greatly reduced.

%%%%%%%%%%%%%%%%%%%%%%%%%%%%%%%%%%%
\subsection{Final State}
%%%%%%%%%%%%%%%%%%%%%%%%%%%%%%%%%%%

\begin{figure}
 \begin{center}
    \includegraphics[angle=-90,width=7.7cm]{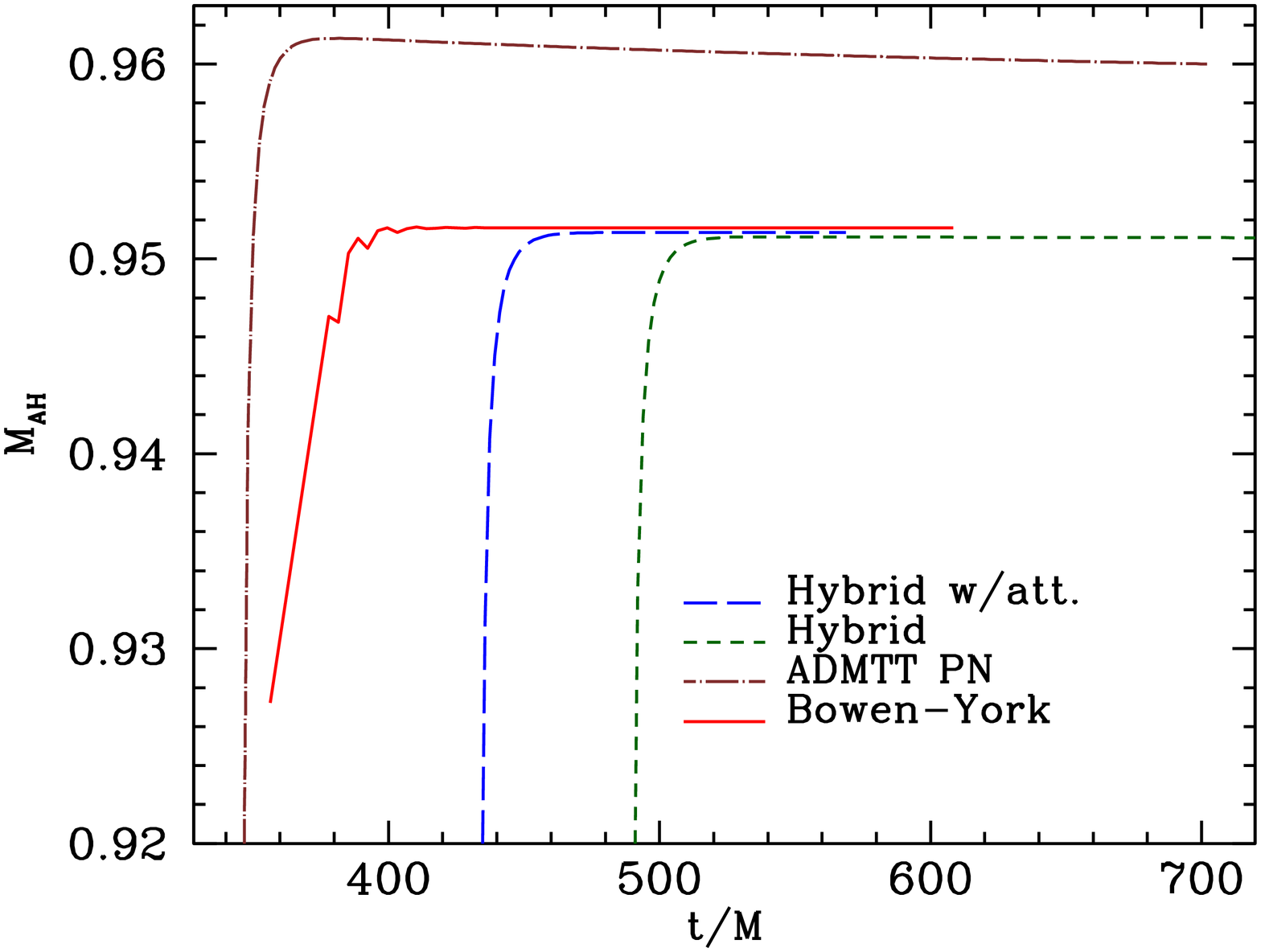}
    \includegraphics[angle=-90,width=7.7cm]{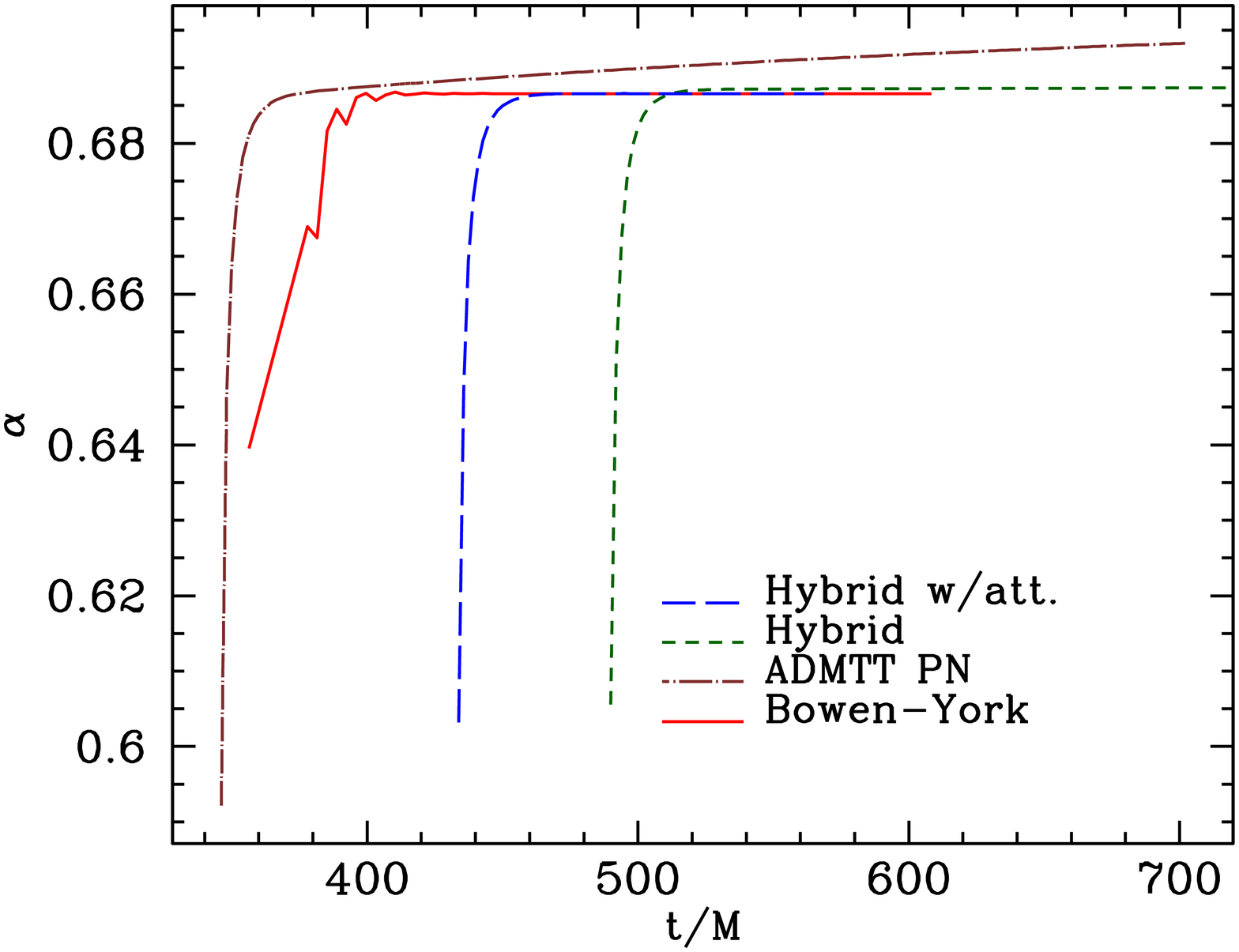}
 \end{center}
 \caption{The apparent horizon mass and the dimensionless spin of the final 
black hole resulting from the evolutions of the Bowen-York, ADM-TT PN, 
Hybrid and Hybrid with attenuation initial data.}
\label{fig:final}
\end{figure}

Finally, we would like to comment on another advantage of the Hybrid data 
over the PN data. As Fig.~\ref{fig:final} shows, the final 
state of the Hybrid data is much closer to the Bowen-York data than 
the PN one. Additionally, the post-merger horizon mass shows a significantly
larger drift over time for the PN data than for the Hybrid data.

%%%%%%%%%%%%%%%%%%%%%%%%%%%%%%%%%%%%%%%%%%%%%%%%%%%%%%%%%%%%%%%%%%%%%%%%
\section{Conclusion}\label{sec:CON}
%%%%%%%%%%%%%%%%%%%%%%%%%%%%%%%%%%%%%%%%%%%%%%%%%%%%%%%%%%%%%%%%%%%%%%%%

Motivated by the benefits that the PN initial data could
bring in bridging the fields of numerical and analytical relativity,
we have introduced a hybrid approach to the initial data evaluation. We
were able to significantly reduce the unphysical horizon mass loss,
as well as reduce the
eccentricity present in the PN data in ADM-TT coordinates,
both undesired features previously reported in \cite{Kelly:2009js}. The
method consists of solving for the traditional Bowen-York puncture data
conformal factor
and using this conformal factor to rescale the ADM canonical
quantities as they appear in the PN data.  The high
eccentricity still present in the data was reduced by the use of an
attenuation function in the black-hole inner zone, effectively
decreasing the contributions of the higher-order PN terms
and the transverse-traceless part of the metric to the canonical
conjugate momentum and three-metric, respectively.

We plan to continue our studies with two lines of research. 
First, as a short-term project, we would like to reduce the eccentricity
of the hybrid data even further by exploring different attenuation functions
and by the iterative eccentricity reduction procedure described in
Pfeiffer \etal~\cite{Pfeiffer:2007yz}.
We plan to use larger orbital separations to facilitate this modeling
and consequently the eccentricity reduction.

The second plan of research, a long-term one, is two-fold. 
On the analytic side, we would like to explore alternative analytic 
approximations to the metric and extrinsic curvature in the binary
black-hole inner zones, where the point-particle approximation breaks down.
Our plan is to match the PN metric to a perturbed, boosted
Schwarzschild
black hole, a la Yunes \etal~\cite{Yunes:2006iw}, with a trumpet
topology in quasi-isotropic
coordinates~\cite{Hannam:2008sg,Hannam:2009ib,
Immerman:2009ns}; this should provide an analytic solution
with smaller constraint violations.
On the numerical side, we will implement a generic constraint solver
for conformally curved data.
By adopting a constraint-satisfying conformally curved metric we hope to better encode
the past history of the binary and provide an initial data free of junk 
radiation.

%%%%%%%%%%%%%%%%%%%%%%%%%%%%%%%%%%%%%%%%%%%%%%%%%%%%%%%%%%%%%%%%%%%%%%%%
\section*{Acknowledgments}
%%%%%%%%%%%%%%%%%%%%%%%%%%%%%%%%%%%%%%%%%%%%%%%%%%%%%%%%%%%%%%%%%%%%%%%%

We gratefully acknowledge the NSF for financial support from Grants
No. PHY-0722315, No. PHY-0653303, No. PHY-0714388, No. PHY-0722703,
No. DMS-0820923, No. PHY-0929114, No. PHY-0969855, No. PHY-0903782,
No. CDI-1028087; and NASA for financial support from NASA Grants
No. 07-ATFP07-0158 and No. HST-AR-11763. The authors acknowledge 
the Texas Advanced Computing Center (TACC) at The University of Texas at Austin 
for providing HPC resources that have contributed to the research results 
reported within this paper. URL: http://www.tacc.utexas.edu.

%%%%%%%%%%%%%%%%%%%%%%%%%%%%%%%%%%%%%%%%%%%%%%%%%%%%%%%%%%%%%%%%%%%%%%%%
\section*{References}
%%%%%%%%%%%%%%%%%%%%%%%%%%%%%%%%%%%%%%%%%%%%%%%%%%%%%%%%%%%%%%%%%%%%%%%%

\bibliographystyle{iopart-num}

\bibliography{../bibtex/references}

\end{document}

%% file: definitions.tex
\def\bi{\begin{itemize}}
\def\ei{\end{itemize}}
\def\be{\begin{equation}}
\def\ee{\end{equation}}
\def\bean{\begin{eqnarray*}}
\def\eean{\end{eqnarray*}}

\newcommand{\bie}{\begin{itemize}}
\newcommand{\eie}{\end{itemize}}
\newcommand{\ben}{\begin{enumerate}}
\newcommand{\een}{\end{enumerate}}

\newcommand{\beq}{\begin{equation}}
\newcommand{\eeq}{\end{equation}}
\newcommand{\bea}{\begin{eqnarray}}
\newcommand{\eea}{\end{eqnarray}}
\newcommand{\ba}{\begin{array}}
\newcommand{\ea}{\end{array}}

\newcommand{\fr}{\frac}

\newcommand{\ga}{\gamma}
\newcommand{\de}{\delta}
\newcommand{\ep}{\epsilon}

\newcommand{\De}{\Delta}

\newcommand{\Om}{\Omega}

\newcommand{\lb}{\left(}
\newcommand{\rb}{\right)}